# Big Deal cancellations and scholarly publishing: Insights from faculty and graduate student interviews

Madelaine Hare[1], Philippe Mongeon[2], Samuel Cassady[3], & Catherine Johnson[4]

[1]maddie.hare@dal.ca
https://orcid.org/0000-0002-2123-9518
Department of Information Science, Dalhousie University

[2]pmongeon@dal.ca
https://orcid.org/0000-0003-1021-059X
Department of Information Science, Dalhousie University, Centre interuniversitaire de recherche sur la science et la technologie (CIRST), Université du Québec à Montréal

[3]scassady@uwo.ca
https://orcid.org/0000-0001-7033-3458
Western Libraries, University of Western Ontario

[4]cjohn24@uwo.ca
https://orcid.org/0009-0002-5328-9437
Faculty of Information and Media Studies, University of Western Ontario

[*]Corresponding author: maddie.hare@dal.ca

## Keywords

## Abstract

Big Deal cancellations are increasingly undertaken by academic librarians faced with rising subscription costs and shrinking collections budgets. While past research has focused on librarians' decision-making processes and communication strategies, this study aims to understand the perspectives and experiences of faculty and graduate students with Big Deal cancellations through interviews at three medium-sized Canadian institutions. It considers cancellations as a collaborative process of information exchange, rather than a top-down process. This study's findings can inform how and regard cancellation projects can be undertaken with enhanced understandings of their lived realities of Big Deals and the current state of scholarly publishing.

## 1. Introduction

The term 'Big Deals' was coined by Kenneth Frazier in 2001 after electronic journal (e-journal) packages emerged in the mid-1990s (Frazier, 2001). Catalyzed by the rise of digital publishing, commercial publishers offered academic libraries the option to purchase subscriptions to bundles of journals under multi-year contracts, seemingly as a pragmatic cost-saving measure that expanded their collections faster than individual purchasing would allow (Mongeon et al., 2021). Big Deals also facilitated the enhanced efficiency of electronic resource management by reducing the number of licenses and access platforms managed by librarians (Glasser, 2013). Many libraries across the globe took advantage of the availability of Big Deals, while others opted to continue purchasing journal subscriptions à la carte (Bergstrom et al., 2014). Libraries that did opt into these licensing agreements were confronted in the following years with rising subscription costs that exceeded annual library purchasing budgets, prompting the need to cancel these Big Deals in the name of financial survival (CARL, 2016; Hoeve, 2019).

The cancellation or "unbundling" of Big Deals are usually not clearly delineated activities involving the simple termination or non-renewal of a subscription contract. They are complex processes often triggered by strained budgets requiring cost-cutting measures (Ivanov et al., 2020). Librarians will usually weigh the value of journals (typically predicated on diverse measures of usage) against their cost (Jurczyk & Jacobs, 2014; Botero et al., 2008; Lemley & Li, 2015). This process is generally complex and lengthy, involving data-driven decision-making supplemented by qualitative insights from faculty as to what resources they find the most necessary in their research and teaching (Enoch & Harker, 2015; Hardy et al., 2016; Gagnon, 2017; Hoeve, 2019; McLean & Ladd, 2021; Ambi et al., 2016). Librarians will often work in conjunction with senior library or institutional administration who govern the process or have the final say on journal cuts. Strategic communications related to journal cancellations at the library will be undertaken by librarians either throughout or after the decision-making process, often leveraging the knowledge of liaison librarians and their relationships with faculty to do this tactfully and effectively (Johnson & Cassady, 2024; McLean et al., 2021).

Journal subscriptions and cancellation projects exist in the context of the current scholarly publishing system. New developments such as read-and-publish deals and open access models factor into how academics navigate the academic publishing space and influence their understanding and acceptance of Big Deal cancellations. It has been found that faculty and graduate students are highly interested in scholarly publishing, but are lacking in knowledge, particularly about subscription and publishing costs (Owens & Manolovitz, 2021; Schiavo, 2024; Welhouse & Boock, 2023). Levels of awareness and knowledge about scholarly publishing influence faculty and graduate students' publishing behaviours, as well as their perception of and reactions towards cancellation projects (Lusk et al., 2023; Webber & Wiegand, 2022).

Cancellation projects occurring at higher education institutions in the past several years have generated varied reactions from faculty and students. Cuts at the University of Ottawa in 2016, for example, elicited concerns related to researchers' ability to effectively undertake their work and the reputation of the institution (Mills, 2016). Journal cancellations at Memorial University of Newfoundland also generated apprehensions about graduate students' capacity to complete their research and speculation about how losses in access to resources might impact the recruitment of future students (Howells, 2015). At the University of Calgary, while faculty's shock resulted in panic around the cancellations, cuts were conveyed by the administration as a step in the right direction to purchase resources singularly and purposefully to address faculty interests; alternative access modes were also emphasized (Strasser, 2017). McLean et al. (2021) note that communications campaigns to faculty and students can make crucial differences in how cancellation reviews and decisions are understood and supported. Indeed, responses often parallel the campus community's awareness of challenges in scholarly publishing and their economic implications for academic libraries (Olsson et al., 2020).

## 1.1 Research objectives

In the past two decades, librarians, as service-oriented practitioners, have concerned themselves with how best to consult faculty and students in cancellation decisions, educate them about the need for and process of cancellations, and deliver messaging in tactical ways that both inform and assure faculty that their academic activities will not be adversely affected by cancellations (McLean et al., 2021). This exploratory study aims to provide a nuanced understanding of faculty and graduate students' awareness, knowledge, perceptions, and experiences related to scholarly publishing and Big Deal cancellation projects at their respective institutions. It considers Big Deal cancellations as a collaborative process of information exchange, rather than a top-down process. This investigation is part of a larger project (Johnson & Cassady, 2024; Cassady et al., 2025) examining Big Deal cancellation processes at Canadian universities.

# 2. Methods

## 2.1 Data collection and processing

### 2.1.1 Recruitment

Data were collected through semi-structured interviews in English or French with 6 university faculty and 7 PhD students (some who also work as lecturers or part-time faculty) at three medium-sized Canadian universities between 2020 and 2022 (Table 1). Faculty and graduate students were invited to participate in interviews through email.

Table 1. Interviewee summary

| University | Interviewed faculty members | Interviewed PhD students | Total |
|---|---|---|---|
| Lyra U | 2 | 3 | 5 |
| Orion U | 1 | 1 | 2 |
| Andromeda U | 3 | 3 | 6 |
| **Total** | **6** | **7** | **13** |

The names of higher education institutions and interview participants have been anonymized, and all publisher names have been omitted. Note that participant codenames indicate their affiliation (first letter: L, O, or A) and role (F represents a faculty member, and G a graduate student), for example: participant L3G is a graduate student at Lyra U.

### 2.1.2 Interviews

Faculty members and graduate students (PhDs) who consented to participate in our study and have their audio recorded engaged in 30 minutes to one-hour interviews over Zoom. Two to four members of the research team conducted interview (see Appendix A). Recordings were transcribed using the online transcription service Scribie (https://scribie.com/). Six interviews were transcribed by French-speaking research assistants. Three of these were transcribed into English by research assistants and three were translated into English by DeepL software (https://www.deepl.com/en/translator) and checked for accuracy by the French-speaking research assistants.

## 2.2 Data analysis

### 2.2.1 Codebook development and coding process

Interviews were coded by at least two members of the research team in Excel. A codebook was developed through close reading of transcripts and in conversation with the research team. Recurring patterns were consolidated for consistency until consensus was reached on nine broad themes: Participant, Knowledge of journal crisis, Access, Connection to library, Cancellation project, Determining journal value, Cost, Open access, and Current publishing model, encompassing 39 sub-themes (Appendix B), which were used to code transcripts.

### 2.2.2 Thematic analysis

Thematic analysis of the coded data was conducted to identify recurrent topics and concepts discussed by interview participants (Urquhart et al., 2010). Clarke and Braun (2016) explain the utility of thematic analysis for "experiential research which seeks to understand what participants think, feel, and do", which can be used for inductive analysis on small, heterogenous datasets, and capturing both manifest and latent meaning (p. 297). This analysis provides a nuanced understanding of the awareness, knowledge, perception, and experiences of faculty and graduate students with Big Deal cancellations at their institution.

### 2.2.3 Statistical summary

A descriptive statistical summary of the responses by each participant for each theme was generated using R Statistical Software (v4.2.1; R Core Team 2022) and is presented in Table 2.
It shows the respondent occurrences for each theme (one for each interview participant), showing that every theme has all participants refer to it. Access had the highest average response rate per participant, followed by Determining Journal Value and Current Publishing Model. Interestingly, Access, Current Publishing Model, and Determining Journal Value had the highest standard deviation (sd), meaning some participants were likely to mention them at high rates and others very low. This is indicated through the minimum and maximum responses, which shows that Access has the highest standard deviation and was mentioned 30 times by one interviewee, and only 4 by another. The lower response rate and low standard deviation of Cost, Open Access, and Knowledge of Journal Crisis reveals that participants may consider these themes of either lower importance or possess less awareness or knowledge about them.

Table 2. Statistical summary of responses to each code

| Code | n | mean | sd | var | min | q1 | median | q3 | max |
|---|---|---|---|---|---|---|---|---|---|
| Access | 13 | 14.769 | 8.852 | 78.359 | 4 | 9 | 11 | 21 | 30 |
| Determining journal value | 13 | 11.000 | 4.983 | 24.833 | 4 | 7 | 10 | 15 | 21 |
| Current publishing model | 13 | 10.769 | 4.106 | 16.859 | 6 | 8 | 10 | 13 | 20 |
| Connection to library | 13 | 9.615 | 4.350 | 18.923 | 5 | 6 | 10 | 11 | 20 |
| Cancellation project | 13 | 8.308 | 4.733 | 22.397 | 2 | 5 | 8 | 10 | 20 |
| Cost | 13 | 7.538 | 4.352 | 18.936 | 2 | 4 | 8 | 10 | 15 |
| Open access | 13 | 5.231 | 2.713 | 7.359 | 1 | 4 | 5 | 6 | 10 |
| Knowledge of journal crisis | 13 | 4.385 | 3.280 | 10.756 | 0 | 2 | 4 | 6 | 12 |
| Participant | 13 | 3.462 | 3.126 | 9.769 | 1 | 1 | 3 | 5 | 12 |

Note: The Participant theme (comprising the sub-themes Position, Department or faculty, Years of experience, and Area of research) was used to identify the personal characteristics of interviewees and provide context to coded responses. While not presented in this analysis to protect the privacy of participants, this data aided with the interpretation of responses for each theme.

## 3. Findings

This section first presents a statistical summary of the results of interview data to illustrate at a high level the main themes that participants engaged with. Findings are organized in sections by each of the eight broader themes, and contextualized with participant responses to provide granular insights, illustrating faculty and students' perceptions of and experiences with each one. The most and illuminating and representative responses were purposefully selected to reflect strong patterns in the data and highlight common opinion, or divergence from it (Lingard, 2019).

### 3.1 Knowledge of journal crisis
#### 3.1.1 Awareness and knowledge of the journal crisis
Faculty and graduate students were asked about their knowledge of the lucrative business models prevalent in academic publishing, an important context for understanding the necessity and implications of Big Deal cancellations. Interviewee knowledge ranged from minimal understanding: "I must confess that I have no idea how much it costs to access a journal. So, I have no idea what is at stake here" (A5G), to general understanding: "I know that it costs a fortune and that the big publishing houses are very voracious and that they have an extremely profitable business model where we work for them for free and then after that they turn around and charge ridiculous prices to libraries" (A4G). A few participants were able to distinguish between individual and bundle subscriptions but were uncertain of how they worked (L1F; L2F; L5G). Some reported no knowledge about the issue, noting it was not on their radar (L4G).

Some Faculty felt that awareness of this issue by colleagues was generally low: "I don't know how many of my colleagues know that there are bulk subscriptions" (L1F). Another participant mentioned:

> "I'm a little disappointed, with just sort of, like, the knowledge of how academics and universities work…I have realized over the years that I am the only one that is really paying attention to a lot of these things" (O2G).

Some participants' awareness of the crisis was limited to the financial strains of the university:

> "No, I have a very vague idea, but I know that they require, they acquire subscriptions and handle journals through that way. I know that we have an embedded librarian in our college who consults with our faculty and faculty in other colleges, obviously about areas of research interest that require support from library services, but the business model and how they go about it, they get a budget from the central university. And that's about the extent of my knowledge" (O1F).

However, they were fuzzy on the specifics: "I know I don't know the prices. I have a general idea, but I also know that there are package deals because you subscribe with Routledge, but you know, the more you take, it's like anything, the more discount you get. But after that, you know, let's say Routledge, it's a half million bill, it's a $250,000 bill? Now I have no idea" (A6F).

Faculty and students were also somewhat aware of other university's responses to the journal crisis (A6F; A5G), particularly with instances that elicited more publicity, such as the University of California's boycott of certain publishers and consequent cancellations (O1F; O2G). These sorts of larger initiatives interviewee O2G hoped would compel other institutions to similar action: "I think what universities are doing right now in terms of cancelling a lot of these subscriptions will hopefully help". In this way, the size and reputation institutions pushing back against price gouging was also deemed significant for inspiring others to follow suit.

### 3.1.2 How faculty are informed of the journal crisis

Information channels from which faculty derived their knowledge of the journal crisis included their respective institutions, higher education magazines, social media platforms (namely Twitter, now X), and conversations occurring within disciplines. This information sharing was typically "nothing really substantial, beyond what was kind of going on, more on Twitter, or social media discussions, with people in other places within the university, so the librarians, as well as just the broader conversation in general, but nothing super localized" (O2G).

Faculty also tended get information from librarians: "Anyway, what I did is I talk to them and that's how I learn the gossip," or at the library, "Sometimes my students used to work at the library, and they would overhear things" (L2F). Professional or student connections were also a source of information:

> "I was part of the original group of eCampusOntario Fellows. So, we did a whole thing where we critiqued very heavily the business model of journals like, 'Why are we paying for this stuff in the first place when it's public dollars that are funding everything that's being done?'" (L3G).

A PhD student noted that "[…] I had the opportunity for almost two years to attend departmental assemblies. I never saw the issue raised among the faculty" (A5G). This indicates how immersion and reliance on certain systems can result in a lack of awareness, often observed more clearly by those external to them.

## 3.2 Current publishing model
### 3.2.1 Faculty and graduate students' attitudes towards the scholarly publishing system
Every interviewee viewed the scholarly publishing system as, simply, absurd, referring to it as "a parasitic system that makes money off of a community that is publicly funded" (A6F), "appalling and shameful" and "very greedy" (A1F), a "cash grab" (L2F), "extremely messed up" (L3G), and noting "the cost of these journal subscriptions is astronomical and frankly stupid" (O2G). One felt "gobsmacked at the highway robbery that seemed to be...evident in the cost" (O1F), and another considered "I guess I'm trying to think of any other industry where someone is paying you to then go and do free labor for another for-profit company…it's insane" (O2G). L1F pointed out that "academic publishers have some of the highest profit margins of any field". O1F noted that "the publishers are in essence, biting the hand that feeds them". The notion of "theft" (A1F) and the idea of "double-dipping", or publishers profiting multiple times from the same work were mentioned: "Society pays four times for our work. We find that a bit much" (A2F).

Another element faculty and graduate students took issue with was the perceived lack of added value despite high fees. The rise of digital publishing was thought to reduce or eliminate costs and so participants felt that they should be receiving higher quality review and editorial services than they currently are. Many commented on how most of the work was done by academics from universities: "The editors of the journals work at universities. The people who publish the stuff work at universities. The people who do the peer review work at universities. All of those people are paid by the universities, they're doing it to build their CV probably" (L3G). A5G summarized the current state succinctly: "A system of professors who are paid by universities to produce science that will be validated by a publication that will have to be bought back by the university. It is nonsense".

Further, interviewees felt that the current system perpetuated unreasonable rates of production: "There may not be an overproduction, but there is a hyperproduction, all the same, of scientific references. Then it has to find outlets" (A3G). One viewed this as the consequence of a capitalist system but was not entirely against for-profit publishing, but felt moderating it was vital to ensure it was not adversely capitalized on: "I'm not against… commercial publishers who make a little profit…It's all a *question of proportion*" (A2F). This high rate of production was thought to generate other questionable practices in academic publishing:

> "We compare with what Bourdieu wrote…It is perhaps this in fact that motivates many to quote so much in our articles. To spin the wheel a little, to quote to be quoted. So, in relation to that I am quite critical. To what extent it is really connected with the business model... I think it's probably linked a little bit" (A3G).

A2F stated, "Well, I would say that we can also hold up a mirror to ourselves, with the tendency to overpublish that is developing, so the monster that is in front of us. We have partly created it.

And even if we are aware of it, I am not very sure that we are doing much to stop it". In essence, the publishing system was described as a positive feedback loop because publishers ultimately "create dependency" (A6F).

### 3.2.2 Ways of shifting the needle

The ability to create change was viewed by many as difficult due to its entrenchment in scholarly communication, though achievable: "And there are so many problems…with obviously a system that is well installed. So, getting the system to move is complex" (A5G). Some felt that change needed to be top-down and involve legislation to regulate the publishing industry: "They're big publishers. You know, it's a very, very, very powerful oligopoly. And they still control quite a bit of what's done. So, unless, you know, there's a law, something where the federal government gets involved" (A6F). Another call for government intervention was issued by L4G: "I think it should be government. I think it can definitely be a collective, collaborative approach. And you're going to get economic inequities there…If we want research to matter to people, we need it to be accessible to them". Some participants felt publishing should be operated by academic institutions, such as universities or funding agencies (L3G; L2F).

Among those that considered change an easier reality, A6F stated:

> "That system shouldn't even really exist, because it's a negation of the very origins of what we do… It's pretty much one of the easiest systems in the world to change... When you look at who's doing the work and everything, it's us. Take away those people, we're still here. We do 90% of the work."

They described buying back publicly funded research which remained inaccessible as "Kafkaesque" (A6F).

One graduate student was aware their institution mandated open access:

> "I know that there is a global reflection on accessibility. I know that now Andromeda U is pushing…I know that it is their policy, so I know that there are reflections in this direction" (A5G).

Participant A6F felt strongly that change was not to be negotiated with publishers but seized by authors "You don't discuss the terms of his demise with him, you eliminate him and then say, 'Look, you've still made good cash for all these years.'" This was because, A5G felt, publishers intentionally make their systems unclear and difficult to disentangle: "And I really have the impression…that they are quite comfortable with this opacity".

O2G suggested that the impact of individual action was limited:

> "There are individual actions that people can take by no longer reviewing for certain journals or choosing not to publish in certain journals and that sort of thing. But there's only so far that individual action can go"

A2F advised that change should not be anticipated to occur overnight, especially when taking a humanistic approach to change:

> "I think you have to have a minimum of patience. Of course, when you have a whole editorial team, you don't necessarily want to throw them out overnight, especially since they're people you know. And so, if it has to take…fifteen years for this transformation to be done, when technically it could have been done in two years, well if it takes into account the people, I'm not hostile."

In addition to reworking the existing publishing model, faculty discussed the difficulty in changing the behaviours of academics publishing within those models. Early career researchers were thought to be in more vulnerable positions and so one participant felt it was the responsibility of established faculty to create change:

> "…it has to come from what you called the elite, actually the older ones…people who are at a career level where they don't depend on publications at all. I wouldn't expect it to be associations of doctoral students who militate for this because it would tend to be dangerous" (A2F).

L3G felt the main problem would still be convincing faculty to change their habits due to the larger systems of evaluation in place and the need for academic survival: "The problem is that faculty want tenure at university…a lot of the systems that have been developed support the notion of keeping your publications in these really established journals, with high impact factors, that are behind a paywall". A6F had noticed changes in how quantitative evaluation was conducted for evaluation purposes, however:

> "We in the department have started to try to think the same, I see it in the aggregations, in the promotion files, we don't look that much at the impact factors, it's not important. It's, 'Have you published.' Then if you have, in your field, then you know, we ask a lot more questions because we don't want to start playing that game of saying well, because when you start quantifying everything, you know, I mean, there's no end to it. So, we try to resist in our own evaluations".

### 3.3 Cost
#### 3.3.1 Awareness of financial factors necessitating cancellations
Most faculty and graduate students were aware that journal subscriptions were expensive but were speculative about exact costs. Some had investigated journal costs:

> "I have a general idea because sometimes… I'll come across the subscription prices…You know sometimes you go "Oh my god". You've got $1000, and you've got $5-10,000. And then the thing is, not all journals put the prices up."
> (A6F)

This obscurity was observed by others: "If I go to Communication and Organization, I look at the beginning of the premium: 'To subscribe, please contact...' They don't even give the price" (A5G). Many were unaware of their library's budgets, even when made available to them (O1F).

A few interviewees were aware of budget cuts at their institution that subsequently affected their ability to access journals (O1F; O2G). One faculty member recognized that their awareness (brought about by a communicative Dean of Libraries) might not be typical: "We were more aware of and probably cared more about the budget than maybe the ordinary professor would" (O1F). They continued

> "I hear them blaming the budget and whether they understand that the costs are extraordinary because…at least, partly laid at the feet of the publishing houses, as well as reductions in university budgets. I'm not sure if they understand where the causes for the additional expenses are coming from, but I think they understand that there was essentially no choice" (O1F).

Some participants were empathetic to the library's predicament with Big Deal cancellations, commenting that budget cuts were not an indication of the university's devaluation of their work, but rather the exercise of effective financial management: "And so, I don't see those lack of connections to those articles as being not supportive of my research, I see that as very much a budget, and just the cost-benefit analysis" (O2G).

Some were ambivalent; acknowledging the tension between financial feasibility and the responsibilities of an institution: "I know it's very, very expensive which is why I tread carefully when I criticize things because we're limited with funding" (L5G).

### 3.4 Cancellation projects

When asked specifically about cancellation projects, faculty and graduate students expressed a variety of opinions regarding how they were consulted and/or how cancellations were communicated. Discrepancies resulting from gaps in these processes and any resulting changes in access had strong bearing on the attitudes of faculty and graduate students towards the cancellations.

#### 3.4.1 How cancellations were communicated

Methods of communicating about cancellation projects varied. Information sharing via word of mouth occurred more pre-pandemic (L2F). Cancellation notices were delivered via email, institutional websites, social media, and information sessions. Library leadership and liaisons were lauded for being particularly effective at communicating with faculty: "Our Dean of Libraries really did do a very thorough job in terms of trying to inform us as to the 'whys' of what was going on, and because we do have a system of embedded librarians, it helps to convey the information" (O1F). Similarly, participant A2F found their Director of Libraries convincing:

> "Then his presentation was very clear as to the reasons, the calculations that had led to it, the way to mitigate precisely the effects by saying, 'Whatever you ask yourself, we'll order piecemeal and then re-evaluate the costs.' All this had been extremely well thought out and creating new power relationships seemed to me at that time necessary" (A2F).

Faculty in turn attempted to support such efforts: "When she would come to me as the Dean of the college, I always made the time to get her onto the agenda for any meetings…I think it is critically important" (O1F). Others found that communications were lacking: "we did have some meetings with our library services and that was never something that came up... it wasn't something that was a direct message of, Hey, you're no longer going to have access" (L3G).

Faculty and graduate students also learned of cancellations through happenstance, such as through their research process when they could not access certain publications (L4G). Participant O2G felt that faculty were bound to discover these cancellations if they were actively conducting research: "Do they research? I always wonder when people don't realize that they don't have access, it's always something that fascinates me, I don't know how they don't run into that". Some learned of cancellations at their institution through this study:

> "[W]hen this study invitation came out, that was the first I heard that there was stuff that was cancelled. I was like, 'What, really? Maybe that explains why I can't get some stuff'" (L3G).

L5G noted that they only wanted to be informed about cancellations relevant to their field:

> "I wouldn't want to hear about every cancellation, only within the health sciences literature…but that might be difficult to even figure out who needs to hear that. It could be…on the library website, just a web page to say the status, so if there was a decision related to a cancellation, that could be there".

Students felt that they were not in the loop:

> "Most of the times when you do interact with our supervisors… there wasn't a conversation of, 'Hey, are any of you having issues finding sources now with these changes, is it impacting your research and what you're able to get access to?' So that hasn't come up" (L4G).

### 3.4.2 Attitudes towards cancellations
Participants' attitudes about Big Deal cancellations ranged from annoyance to strong objection, though this did not jade their attitude toward the library. Some mentioned their right to access information: "I have a philosophical objection because we are being... It feels almost like being censored when you don't have access to a wide range of journals" (O1F). Some participants, like L2F, expressed an opposition to cancellations:

> "We got a letter some time ago, about a seemingly innocuous thing, about how to serve you better kind of thing…And I told everybody, 'My God, they are going to cut.' That's why they are asking. We are not that stupid, you know. I know

Big Deal cancellations 11

the pattern. I've been around. "We are not cutting, but can you tell us what kind of journals you want?' I said, 'All of them'" (L2F).

Participants largely felt that they lacked agency: "In the end, I have very little recourse to the services of the university" (A3G). L2F felt they lacked organized action: "What do you do, you pick up a shovel and chase them? There is nothing we can do. Nothing. We cannot strike, what can we do?" A6F explained, "I would say probably because we're pretty much all on the same page here, in a sense that there's no one who wants to defend a system like that tooth and nail."

Some faculty attitudes towards journal cancellations were positive as they resisted for-profit publishing: "My reaction would have been, 'Ah, yes, good, we are divesting from [publisher] to some limited extent, good job.' And then I would have carried on" (L1F).

### 3.5 Access
#### 3.5.1 Alternative means of accessing resources
Maintaining access is a key consideration when navigating Big Deal cancellations. Faculty mitigated loss of access in myriad ways with some advantages and disadvantages. Placing an inter-library loan (ILL) request was among the most common steps taken, either revered for its convenience or dislike for its tediousness. O2G noted that a few extra minutes of "online sleuthing" or waiting an extra day to receive an article still saved them a trip to the library, whereas LG3 perceived ILL as a "painful" and time-consuming process. Alternate access methods used by participants were not only influenced by a motivation for access but how data might inform decision-making. One interviewee voiced a commitment to placing ILL requests to provide librarians with data: "If it's…from a journal that is one that I would really like access to, I try to…put in a request through the library itself so that they have that information in their system that researchers are trying to access particular titles" (O2G). A graduate student noted that when pursuing alternative routes of access, they still place requests on an item, "at least logging that so that the libraries have that information because otherwise I realize if the library doesn't know that people need it, they're not going to have that data" (O2G). L5G stated: "I know that that usage gets tracked and I'll try to pick a nursing database to give more hits to something that I recognize, something that I'm trying to support.".

Another common means of accessing information was contacting authors of articles (A5G), which also acted as a networking mechanism: "Actually, the reason that I would maybe want to contact the author is if it is very relevant to my research, and it's a contact that I want to make anyways" (O2G). Some found it more expeditious: "I reached out to the author. I could find her, she's from Australia, I emailed her and someone from her team emailed me less than a day" (L5G). Some relied on their supervisors who are more embedded in the research system, though the case of L3G shows that a researcher's *distance* to a journal at times does not influence their ability to access it: "My thesis supervisor publishes a lot in that journal…and I was like, 'I can't access it', and she couldn't access it either, so she emailed the author. And she's one of the editors of the journal" (L3G). Others outsourced access to their administrative support: "I send her the article that I want and then she sends me a link...I feel like she's making arrangements with other librarians and then there's some kind of pool" (A1F).

Interviewees also used other credentials for institutions to which they are affiliated to gain access to their catalogues (O1F, L5G).

Apart from social connections, participants also relied on information technology outside of their library's online catalogue: search engines such as Google Scholar (L3G) were commonly used, while others went straight to illegal sites like Sci-Hub for its speed and ease of use: "In the worst case, I happened to go through some Russian sites that were of questionable quality. But politically I had absolutely no problem with it, as all this research was done financially and publicly" (A6F). ResearchGate was commonly used, though some noted its inconsistency: "Through ResearchGate…you can get access to some of the ones that have been made accessible…others you have to actually request the individual to give you access to it…it's very inconsistent" (L4G). This irregularity influenced researchers' use of academic libraries and made them cautious about going to it as a first place to search:

> "I will directly access what I need through other paths on the internet. Then the second solution is that I go through the university's proxy to have access to the few remaining periodicals, which are becoming fewer and fewer" (A1F).

The proliferation of preprint servers also provided a new gateway to information, even mitigating larger concerns around access (O2G). O2G noted that "I was really hesitant towards pre-prints when I first started my PhD, I kind of saw them as super sketchy, but I think the use of pre-prints might be changing people's use of articles too".

### 3.5.2 Concerns around a loss of access

Alternatives were often discussed in relation to the amount of time needed to compensate for a loss of access, though some participants held conflicting views. O2G felt that "hunting down articles" took extra time, but "I would say that my research is still plodding ahead at the same speed and that it hasn't really affected me a lot more than just a minor annoyance at this point". Other participants differed: "Now, three-quarters of the time, the article is not there anymore. So, there is a kind of lag. Frankly, it's a pain" (A1F). Some participants accepted the lack of access: "A lot of the time, it's just like, 'I can't get this journal. I'm going to have to do without it" (O1F), even prompting them to consider whether the source was necessary to their work: "I might say, 'Well, do I really need that article?' And I kind of go back to the abstract and then reassess my research question and say, 'Oh... Maybe it's not actually that important'" (O2G). In this case, the one drawback acknowledged was the loss of visibility for the author of the article.

Access concerns were also expressed about groups who would be disproportionately affected, including smaller disciplines, early career researchers, and students. One participant noted that junior faculty, "will have greater difficulty in accessing the various pieces of work that they need in order to get themselves established" (O1F). L2F noted, "I remember the place where I most frequently can't get access to an article that I would like access to, is for undergraduate classes, students are doing research". L1F echoed the same concerns about student access: "I have other ways of accessing journals, students don't". L1F also expressed concern about smaller disciplines: "I would be concerned that I and then presumably other people in very small disciplines at the university would be the ones most likely to lose access to journals that we needed." L3G was concerned with how cancellations would affect students and informed decisions about where to enroll:

> "If you're getting information that's important for the success of some of your students that have a particular database, they need to know before they sign up that they're not going to have access to this stuff, right? Not when they're in the middle of the program."

Acceptance of a lack of access was influenced by the size and financial state of institutions. Participant A5G felt "lucky enough to be at Andromeda U, a huge research university", whereas L3G noted:

> "I was all excited to go to Lyra U because they have all of these tools, and I thought they would have more access to resources… I figured out that a lot of the resources that I couldn't get through Lyra U, I could actually get through [other university] …you'd think that at a larger university, you'd have access to everything."

Concerns about access were also coloured by the *cause* of the lack of access. One participant voiced more support for Big Deal cancellations if "it's to try to change a business model… what this operating model says today also has an impact on citizens who want to have access to science, who want and are interested in science" A5G took issue with having to "pay more for open access" in an already inequitable system of science (A5G). They worried that a lack of access would affect the diversity of research:

> "That's why I think that these are extremely problematic and complex issues, but they are important issues, the accessibility of research. That is to say that cancellation campaigns if it is to try to change the models, put pressure, yes, go" (A5G).

## 3.6 Connection to library
### 3.6.1 Relationships and interaction with the library
There was consensus among interviewees that "The library…is extremely important" (L2F) and highly used, though the nature of relationships to it changed over time: "Libraries are one of my favorite places and I hang out in the library. I just don't get into the stacks anymore. It's faster to go online" (O1F). Relationships also developed through different channels of communication; some arose through interaction on social media sites such as Twitter, others through proximity in professional environments (O2G). Overall, every participant expressed a positive relationship with librarians and appreciation for their service, noting an alignment of professional values and interests: "There is really a very firm professional conscience and... how shall I say it, perfectly compatible with these ideas that I have developed previously that seem important to me" (A2F). L2F empathized with librarians, feeling that they bore the brunt of faculty dissatisfaction with the results of journal cancellations.

Interviewee's interactions with librarians ranged from minimal: "not something that I would say is very, very frequent" (A4G), "have a little bit of that rapport" (L4G) to major: "our Dean of Libraries led a major information series of sessions to let us know about the issues with regards to these bundled subscriptions and the cost of them" (O1F). The nature of outreach ranged from consultations to collaboration. A few participants noted that librarians were embedded within their departments and made efforts to be present and consult with faculty: "She shows up at our faculty council meetings in the college. She consults with our faculty" (O1F). On the frequency of these consultations A2F relayed librarians: "often come to us and say, 'Here's the list of we have to do, for example, we have to make cuts, what do we have to cut?" Or the other way around, effectively. "Do you have any particular purchasing needs?'". Participant A2F also offered an example of the way they saw librarians collaborate with faculty:

> "Our librarian was doing what he called scouting, in fact, it was all the books he was planning to buy. He would send us roughly the reference of the book, and then a sort of back cover… So, it wasn't him who took the time to write something, not having the time to read all the books he had bought. And then it was sent to all the teachers. I found that extremely useful" (A2F).

### 3.6.2 Faculty consultations and involvement in acquisitions

The consensus was that faculty should be consulted, though there was ample trust in librarians' abilities. For instance, one interviewee noted librarians had a good understanding of the important periodicals to attain and so consultation was sought more frequently concerning other acquisitions:

> "It is more frequent for manuals or books and less for periodicals because I think they know or understand what important periodicals are. So, I don't remember that they would have contacted me to find out whether we should subscribe to such and such a periodical" (A4G).

One participant was more skeptical of librarians' knowledge of crucial journals: "They'd have to delve pretty deep, and it's often easier just to sort of get the superficial, which are the ones that are used the most, kind of questions as opposed to that deeper understanding of which are the ones that are critical for a particular discipline" (O1F).

One participant acknowledged that more could be done on the part of faculty and students to get involved with library acquisition:

> "I think in terms of the role of faculty and students asking for that, in my experience, if you say, 'Hey, this is really important,' it gets it on to the librarian's radar and that can be something that they can work towards, and oftentimes they want to help work towards " (O2G).

Some felt they were not afforded ample opportunity to participate: "We have a library committee on which I sit but it doesn't meet often" (A6F). Another faculty member described how the lack of faculty awareness inhibited their ability to get meaningfully involved in acquisition but felt that they should still be consulted: "But to the extent the faculty can give useful feedback, I think they should be part of the conversation" (L1F).

### 3.7 Determining the value of journals

Surveys were a common methods librarians solicited input on cancellation decisions. Most faculty saw them as valuable for deciding what to cancel and how to fill gaps left from cancellations (A3G), as well as for filling in librarians' gaps in domain knowledge (O2G). Some commended librarians for their commitment to consulting faculty:

> "I know that she goes to considerable effort to track people down and hound them a bit to get that information. When I... receive such a request, I do answer." (O1F)

Others felt that surveys were insufficient for elaborating on why certain journals were important to their research (L4G; L5G). L3G conceded that due to the size of universities, individual consultations were not an entirely realistic approach:

> "I think really a survey is the only thing that they could do…. I think that if they tried to interview everybody … before they made a decision, decisions would never get made." (L3G)

Usage statistics (such as downloads) were seen to be the most objective metric to assess the value of journals. L1F felt that citations mattered less than usage because

> "I'm in a field where no journals have particularly high index factors, so things like how frequently these articles are cited, that'll vary a lot by field. And so, I feel like the actual access by people at the institution of the journal is the important one."

They believed that usage statistics must be balanced against the cost of journals: "Obviously, if a journal is, you know, $5 a year, it probably doesn't matter how many people are accessing it, not that any journal is $5 a year" (O2G).

### 3.8 Open access
#### 3.8.1 Attitudes and concerns about the open access model

Attitudes and experiences with open access publishing illuminate how researchers navigate the rapidly changing research system, and how they make decisions about both consuming and producing research. Open access was viewed favourably among participants, though some were happy to support closed access publications if the funds went to support scholarly societies, for example, because they did "useful things" with the costs of subscription and publishing (L1F). O1F reified the importance of open access for lower-income countries but was not hopeful of its promise:

> "I know that I have a number of international colleagues who are in second and third world countries, and open access is a godsend for them so that they can actually access those some of those... I wish everything were free, but...I'd like to have a pet dragon too, so [chuckle]."

Faculty possessed several concerns around predatory (or deceptive) publishing (O1F; L2F). A few participants had fallen into the trap of predatory publishing before (L2F). L5G noted "we're entering these land mines all the time, it's a very messy space…I think we need more education and support in terms of navigating that. But yeah, I would publish open access".

A PhD student emphasized the need for developing academic literacy:

> "Academic Literacy or something like that, in terms of how do you even navigate that? And I feel like all of us as grad students need to really look at that. I started asking the team, there were many authors on this paper, and I said, 'We've been approached to publish this research. Here's the journal. I can't find it as a predatory journal, does anyone have any concerns?'" (L5G).

Open access was considered by some to be old wine in new bottles- that is, the same problem in a different form:

> "And a lot of conversations with another colleague or a collaborator of mine that's actually very anti-open access has been really interesting…I was like, 'My gosh, but what about access to knowledge?' And then hearing his reasons as to why... It's basically just off-loading the cost to someone else. And it's still the publisher that's making off with sacks of money at the end of the day." (O2G)

### 3.8.2 Attitudes towards publishing open access

APCs were identified as a real barrier to publishing, especially for early career researchers. Participant A5G noted

> "It's certainly the paywall system, the open access system, that becomes a pressure, even on a PhD student. It's a doctoral student like me who is going to publish. He's going to think, 'Okay, am I going to put $3,000 to get my publication open access because I know my open access is going to be more cited? I'm going to get my citations up, my impact factor up, whatever, and so it's going to make me more employable down the road'".

O2G noted that when they were a student, their PI covered the costs of OA journals so that the work they published would receive more citations. APCs were also factored into venue decisions:

> "I think for many, especially depending on how much research you do, if you want this to be your career, and you plan to publish a lot, that can rack up very, very quickly… If I've got 15 publications, am I going to pay for all 15?" (L4G).

One graduate student identified the lack of publication fee support for researchers: "I also recognize that the cost that gets put back on to researchers and just cannibalizes research grants is a significant detriment" (O2G). This was made more difficult, as another researcher noted because "federal, SSHRC or FRQ grants at the provincial level, there are now obligations to publish in open access, at least to a limited extent" (A2F). Another faculty member identified the idiosyncrasies of disciplines as a factor impacting the utility of open access:

> "Open access is of course, important and it's very clear that the Tri-Agencies are pushing us towards open access for all of that, but open access is not currently sufficient for the work that I do because a number of the journals that are critical for my field are not truly open access" (O1F).

O2G observed a citation advantage in publishing open access as compared to closed journals but described how they went two pages over their limit for an open access publication and "it cost us 400 pounds sterling" though they ultimately decided that "the data and the information that was in those pages was worth the publication costs for those page charges."

Apart from hefty publishing fees, however, participants were concerned with how open access journals factored into research evaluation. Participant A4G expressed that:

> "they are generally considered second class, and the problem is a question of quality, of impact. They don't appear in the usual leaderboards and so when it comes time for promotion, well they have no weight. So, it's not recommended".

Analogously, A6F felt there was a "survival aspect" to choosing where to publish, noting that though they felt pressure earlier in their career to publish in top journals, publishing open access was "Definitely something I want to do more and more, now that I have more latitude, and I have more freedom". Prestige was considered a significant factor by another faculty member: "When you are a specialist, it's the ones that you know that you would like to publish in this one or that one for this or that type of article, because they have a reputation" (A2F). Others were more focused on finding venues suitable for their work, stating it was the most important criterion to them and opportunity to set an example: "I think that we ought to be, as a leader, walking the talk…but I still think it's more important to publish in the journal that will most address the audience that you want to reach" (O1F).

## 4. Discussion

This section collates the most significant findings from each theme and connects them to practice. Big Deal cancellations as processes of information sharing and exchange are also discussed.

### 4.1 Involvement in the cancellation process

In terms of consultation and participation in cancellation decisions and the process at large, the majority of faculty and graduate students interviewed generally felt that they were adequately consulted and informed by the library, though surveys or inquiries from the library tended to elicit fear about cuts. Several participants desired a better understanding of what data from faculty and students would prove most useful in their decision-making. Some went to lengths to place ILL requests even if they sought or found alternate access routes to provide librarians with data that a resource was in demand (e.g., O2G and L5G). These participants typically had more knowledge about the context of cancellations and the scholarly publishing or library system than their peers. In fact, disparities in knowledge between faculty and graduate students and how this affected the latter's agency in the scholarly publishing system were observed (O2G; L1F). In this regard, some felt that publishers were purposefully opaque (A5G). Indeed, this is an example of hermeneutical injustice, described by Fricker (2007) within a larger framework of epistemic injustice, where interpretive resources are purposefully denied or obfuscated, impacting one's ability to understand their social experience. This further impresses the notion that those who are more informed can better contribute to such processes, making them more collaborative. The need to improve academic literacy (knowledge and competencies around scholarly publishing and navigating academia) was brought up by participant L5G, while O2G proposed mandatory library training for graduate students. Instilling researchers with academic/publishing literacies would not only empower them to better navigate it themselves but also support librarians who would work with faculty with a better understanding of librarians' unique position as arbiters in scholarly communications

Many participants expressed a desire to help librarians and get involved in the process of journal acquisition and cancellation or belief they had a responsibility to get involved and use their voices (O2G). Some felt prohibited because they were unsure if their input was wanted or useful, or felt they lacked opportunities to participate (A6F; L1F). This suggests that increased contact can thus position libraries to improve the knowledge and academic literacies of faculty and graduate students and better empower them to help. Modes of establishing contact and building relationships with faculty and graduate students that seemed to help achieve this include social media, information sessions, and professional societies. Similarly, exploring different methods of soliciting faculty and graduate student input regarding acquisitions and involving them at earlier stages in collections processes can help include them in cancellation decisions and keep them informed about the library's current holdings.

### 4.2 Experiences with cancellations and their effects

How Big Deal cancellations affected access to journals varied: some faculty adapted to finding alternative methods of access through personal channels (e.g. supervisors, colleague, authors), indicating how the journal crisis may emphasize the importance of networking as well as increased a reliance on senior academics and scientific societies to provide access. Notably, mutual support was a common approach to navigating issues of access, suggesting that academic community's may be coordinating across institutions, disciplines, or even countries. This is emphasized by graduate student testimonials that they relied on their supervisor to provide access. However, even here, access was variable: some faculty could not access material published by journals for which they sat on editorial boards or encountered other barriers. Some researchers remained relatively independent, hunting for resources they needed online through places other than their library's catalogue (e.g., Sci-Hub, ResearchGate, Google Scholar, pre-print servers). One participant declared "I was not missing anything. I can't say that I couldn't do my research" (A6F). Some interviewees possessed an informed appreciation for the tension between balancing library budgets and acquiring needed journals with accompanying challenges, while others were unaware of budgetary concerns, the acquisition process, publishing models, or the journal crisis itself.

Further, participants expressed concern about those more vulnerable to the impacts of cancellation decisions: smaller disciplines, early career researchers, and students. In the same light participants also felt that they lacked means of recourse and that post-cancellation surveys about resource needs were neglected (A3G; A2F). The scenario of A6F, where resources were bought back as a result of their feedback, is an example of successful post-cancellation communication, and other faculty seemed to find even just the possibility of buybacks comforting. Librarians can prioritize vulnerable groups in their decision-making and communication of cancellations through more direct and intense consultation, enhanced messaging regarding alternate access route resources, and follow-ups post cancellations to determine the nature and impact of the cancellation's effects. Maintaining efforts after cancellations to engage with faculty may also mitigate impact or backlash after cuts have been made.

### 4.3 Increasing understanding

There may be more buy-in from faculty and graduate students if there is ideological alignment with the cause of cancellations, such as in the case of A5G, who noted that if they knew that cancellation projects were attempts to shift the current model of publishing, they could accept ramifications such as the loss of access to certain resources. Here, opportunities to appeal to and engage with the value systems of faculty and graduate students may be a strategy librarians can take to increase support, or at least understanding, around cancellations. Accordingly, participants resonated with librarians' firm professional consciences (A2F), and often considered how their own choices would affect others working within the system. For example, O2G would contact authors of works they accessed through other means to let them know their work was being read. This shows a socially conscious approach and identifies how faculty established workarounds not just in relation to access but in making sure scientific mechanisms are not disrupted, like credit (citations and visibility). The effects of cancellations are thus influencing how researchers uphold and engage with the norms of science. These findings indicate that there may be increased buy-in from faculty about cancellations if a moral or philosophical alignment can be established with their cause.

### 4.4 Changing the current publishing model

There was resounding consensus among interviewees that change in the current publishing model was needed. Open access was a notable development in scholarly publishing discussed by participants, but perceptions and attitudes varied in regard to its viability for *accessing* research as compared to *publishing* research. Findings show that open access was a well-used and relied upon alternative method of accessing publications. Participants valued openness for this reason though they encountered limitations in the extent of open material. As such, open access may act as a mediating factor in faculty and graduate student concerns around Big Deal cancellations. However, participants' relationship to open access for knowledge dissemination purposes varied greatly. Concerns around research evaluation and venue prestige were repeatedly cited as a key factor in researchers' choices of where to publish and were a major influence on the uptake of open access. Research assessment reform is an intervention point currently under concerted attention from the scholarly community (e.g., Coalition for Advancing Research Assessment (CoARA) (https://coara.eu/), The Declaration on Research Assessment (DORA) (https://sfdora.org/about-dora/); the findings of this study thus reinforce the current research evaluation system as a significant barrier to change in researchers' behaviours, particularly in decisions to publish open access.

Examples of the individual level action participants adopted included no longer reviewing for or choosing not to publish in specific journals. The idea of libraries and funding agencies as publishers was a potential means of reclaiming control and ensuring the low costs of scholarly publishing, proposed by several participants (L3G; L2F). Indeed, in the past few years there has been a great deal of scholarship on the topic, and increased efforts to move in this direction (Bonn & Furlough, 2015; Gwynn & Craft, 2019; Lippincott, 2017).

### 4.5 Cancellations as processes of information sharing and exchange

This study also illuminated dynamics of information sharing and exchange around Big Deal cancellations and faculty and graduate students. These processes were bidirectional between certain sources and actors (e.g., between librarians and faculty and graduate students), and unidirectional amongst others (e.g., faculty reading a university policy on a website). Interview data revealed varying degrees of awareness and knowledge in faculty and graduate students around Big Deal cancellations and scholarly publishing as a result of these processes. Some participants possessed base level awareness of the fact that cancellations were occurring, and why (e.g., L4G), while others with deeper knowledge were more engaged and wanted to participate in the process (e.g., O2G). Information channels cited in this study through which information sharing and exchange about scholarly publishing and Big Deal cancellations occurred include university websites and policies, librarians, colleagues, staff, students, scholarship, higher education magazines, social media, and participants' own experiences in academia. These findings provoke consideration of how these channels can be harnessed to generate more awareness, but also facilitate knowledge building (i.e., the understanding and ability to critically engage and develop agency to influence or enact change) as a bidirectional process in which faculty and graduate students are actively engaged participants. This would reinforce the development of deeper understanding around Big Deal cancellations and scholarly publishing as not just top-down processes but occurring laterally across actors, and through external channels to the library and institution.

## 5. Conclusion

This study presents results from interviews with 13 faculty or graduate students from three medium-sized Canadian universities and is not representative of all perceptions and experiences of faculty members from higher education institutions. It nonetheless provides an idea of the awareness, knowledge, perspectives, and experiences of faculty and graduate students, and offers insight into how Big Deal cancellation projects may be approached by librarians and institutions as more collaborative processes.

The findings revealed that culture of research evaluation and academic tenure and promotion requirements strongly influence faculty and graduate students' publishing behaviour. Ample literature on research assessment reforms exists (see de Rijcke (2023) for an overview of current debates and developments in research assessment), and the Declaration on Research Assessment (DORA) (https://sfdora.org/) has prompted coordinated change through funders, publishers, and institutions, and researchers. However, much is left to be understood about faculty and graduate students' decision-making and behaviours related to open access. Future work could aim to investigate information channels and the process of information exchange between academics about open access and other aspects of scholarly publishing. Additionally, intersections between the rise of open access and preprint servers with Big Deal cancellations could be further explored.

Several interview participants mentioned wanting to participate in Big Deal cancellation processes but not knowing their place in the process or how they could be of most use to their library colleagues. Developing understandings of how faculty and students could support the library in catalyzing change in the current system of scholarly publishing would prove beneficial for improving practices of mutual support.

### 5.3 Concluding remarks

This study has explored faculty and graduate students' perceptions and experiences with Big Deal cancellations. It is hoped that it can foster reflections on the role of faculty and graduate students as not just recipients of information around cancellations but collaborators in the process. As libraries move away from Big Deals and continue to navigate a constantly evolving scholarly publishing landscape, informed and involved faculty and students may feel less blindsided about cancellation decisions, better prepared and equipped to locate and use resources post-cancellation, more empathetic to the position of librarians, and empowered to navigate scholarly publishing and contribute to meaningful change.

## 6.    Acknowledgements

The authors are grateful for the contributions of Toluwase Victor Asubiaro, Asen Ivanov, Nicole Delellis, and Helene Bigras Dutrisac, who contributed to early stages of this study.

We express appreciation to the editors and reviewers of College & Research Libraries for helpful feedback and editorial assistance.

## 7. Funding

We acknowledge the Social Sciences and Humanities Research Council for providing funding for this project.

## 8. Author contributions

M. H. Data curation, Formal analysis, Writing—original draft, Formal analysis, Validation, Writing—original draft, Writing—review and editing

S. C. Conceptualization, Data curation, Funding acquisition, Investigation, Methodology, Project Administration, Software, Resources, Supervision, Writing—review & editing

C. J. Conceptualization, Data curation, Funding acquisition, Investigation, Methodology, Project Administration, Software, Resources, Supervision, Writing—review & editing

P. M. Data curation, Investigation, Methodology, Supervision, Validation, Visualization, Writing—review & editing.

## 9. Conflicts of interest

There are no conflicts of interest in this project.

## 10. Data availability

Interview data reported in this study was anonymized. For purposes of privacy, interview transcripts and coded data will not be made available.

## 11. Ethical approval

Research ethics approval was granted for this project by Western University Non-Medical Research Ethics Board (# 114491).

Howells, L. (2015, December 8). Memorial University to cancel thousands of journal subscriptions. *CBC News.* https://www.cbc.ca/news/canada/newfoundland-labrador/memorial-university-to-cancel-thousands-of-journal-subscriptions-1.3354711

Ivanov, A., Johnson, C.A., & Cassady, S. (2020), "Unbundling practice: the unbundling of big deal journal packages as an information practice", *Journal of Documentation*. https://doi.org/10.1108/JD-09-2019-0187

Johnson, Catherine Anne, and Samuel Cassady. 2024. "Faculty Response to Journal Cancellations." Collection Management 49 (4): 165–85. https://doi.org/doi:10.1080/01462679.2024.2422589.

Jurczyk, E., & Jacobs, P. (2014). What's the big deal? Collection evaluation at the national level. Portal, 14(4), 617–631. Scopus. https://doi.org/10.1353/pla.2014.0029

Lemley, T., & Li, J. (2015). "Big Deal" Journal Subscription Packages: Are They Worth the Cost? Journal of Electronic Resources in Medical Libraries, 12(1), 1–10. https://doi.org/10.1080/15424065.2015.1001959

Lingard, L. (2019). Beyond the default colon: Effective use of quotes in qualitative research. *Perspectives on Medical Education*, *8*(6), 360–364. https://doi.org/10.1007/s40037-019-00550-7

Lippincott, S. K. (2017). Library as publisher: New models of scholarly communication for a new era. *ATG LLC (Media).* http://dx.doi.org/10.3998/mpub.9944345

Lusk, J. T., Jones, K., Ross, A., & Lecat, V. (2023). Insight into Faculty Open Access Perceptions: A Quantitative Analysis among UAE Faculty. *The New Review of Academic Librarianship*, *29*(3), 219–243. https://doi.org/10.1080/13614533.2022.2122853

McLean, J., Dawson, D., & Sorensen, C. (2021). Communicating Collections Cancellations to Campus: A Qualitative Study. *College & Research Libraries, 82*(1), 19. https://doi.org/10.5860/crl.82.1.19

McLean, J., & Ladd, K. (2021). The Buyback Dilemma: How We Developed a Principle-Based, Data-Driven Approach to Unbundling Big Deals. *The Serials Librarian*, *81*(3–4), 295–311. https://doi.org/10.1080/0361526X.2021.2008582

Mills, S. (2016, October 21). University of Ottawa puts thousands of journals on the chopping block. *CBC News*. https://www.cbc.ca/news/canada/ottawa/university-ottawa-library-budget-journal-cuts-1.3815030

Mongeon, P., Siler, K., Archambault, A., Sugimoto, C. R., & Larivière, V. (2021). Collection Development in the Era of Big Deals. College & Research Libraries, 82(2), 219-. https://doi.org/10.5860/crl.82.2.219
Big Deal cancellations 25

Olsson, L., Lindelöw, C. H., Österlund, L., & Jakobsson, F. (2020). Swedish researchers' responses to the cancellation of the big deal with Elsevier. Insights the UKSG Journal, 33(1). https://doi.org/10.1629/UKSG.521

Owens, E. E., & Manolovitz, T. (2021). Scholarly Communication Outside the R1: Measuring Faculty and Graduate Student Knowledge and Interest at a Doctoral/Professional University. *Journal of Librarianship and Scholarly Communication*, *9*(1), eP2413-. https://doi.org/10.7710/2162-3309.2413

R Core Team. (2022). *R: A language and environment for statistical computing.* R Foundation for Statistical Computing, Vienna, Austria. https://www.R-project.org/

Schiavo, J. H. (2024). Knowledge, Attitudes, and Practices of Health Sciences Faculty Towards Scholarly Open Access and Predatory Publishing. *Medical Reference Services Quarterly*, *43*(3), 243–261. https://doi.org/10.1080/02763869.2024.2373019

Strasser, S. (2017, January 16). The rising price of knowledge: University of Calgary cuts 1,600 academic resources. *The Gauntlet*. https://thegauntlet.ca/2017/01/17/the-rising-price-of-knowledge-university-of-calgary-cuts-1600-academic-resources/

Urquhart, C., Lehmann, H., & Myers, M. D. (2010). Putting the 'theory' back into grounded theory: guidelines for grounded theory studies in information systems. *Information systems journal, 20*(4), 357-381. https://doi.org/10.1111/j.1365-2575.2009.00328.x

Webber, N., & Wiegand, S. (2022). A Multidisciplinary Study of Faculty Knowledge and Attitudes Regarding Predatory Publishing. *Journal of Librarianship and Scholarly Communication*, *10*(1). https://doi.org/10.31274/jlsc.13011

Welhouse, Z., & Boock, M. (2023). Faculty attitudes toward Open Access publishing: Library-led action and outreach. *BiD (Barcelona, Spain)*, *51*. https://doi.org/10.1344/BID2023.51.11
# Appendices
## Appendix A: Interview Questions

1. Before we begin, could you tell us about your area of study/research/teaching, your professional status (or where you are in your doctoral journey), and tell us how long you have been working at __?
2. A) Do you use periodicals in your work? How important are they to your research/teaching? Are there any journals that you feel are essential to your work? Does the university provide access to these periodicals? How important is the timeliness of the periodicals you use – do you need access to the most recent works or is it acceptable for you to have access to content 6 months or a year after publication? Have you ever been unable to access an item you needed?

Big Deal cancellations 26

B) If you can't access a periodical you need, what do you do? Could you tell us about a time when you had to use a means other than your library to access an article? vs. Do you use print periodicals? (When was the last time you used the in-person library and print resources?)
3. Have you ever asked the library to subscribe to periodicals you need? Do you consult librarians about the resources you need? Is there a particular librarian or librarians you regularly consult regarding access to resources? What is your reaction when librarians call upon professors to identify important periodicals? Do you think professors and doctoral students should be involved in the selection or cancellation of resources?
4. If you were to develop a list of criteria (e.g., cost, use, faculty use, references/citations, librarian expertise, etc.) to follow in selecting titles for cancellation, what criteria would you consider? as the most important? The least important? How convinced are you of the ability of librarians to manage library collections? b. Have you ever complained about unsubscribing from one or more periodicals?
5. Do you have any idea of the cost of e-journal subscriptions at your university? Have these costs been the subject of controversy within your institution? Are you aware of open access periodicals? Do you think they are sufficient for your research? Have you ever published in an open access journal? Do you have any other ideas on ways to reduce the cost of access to scholarly journals?
6. What concerns you most about the consequences of unsubscribing from a large number of periodicals? (Ask about hopes/fears if necessary - what is the best/worst case scenario?)
7. What do you generally think of the current scholarly communication model where publishers get content, peer review and editorial work for free, then sell libraries access to these journals?

## Appendix B: Themes and sub-themes

| Themes and sub-themes | Code occurrences |
|---|---|
| **Participant**  Position  Department or faculty  Years of experience  Area of research | |
| *Subtotal* | *45* |
| **Knowledge of Journal Crisis**  Knowledge of journal acquisition process  Knowledge of library concerns regarding business model  Knowledge of other university's response to journal crisis  Sources of information about journal crisis | |
| *Subtotal* | *57* |
| **Access**  Alternative methods to access journals (including both illegal and legal methods)  Examples of accessing journal the library does not have | |
| Concerns about access (positive or negative) | |
| Concerns around access-currency | |
| Concerns around access-cost | |
| Use of print journals | |
| Impact of COVID | |
| Method of searching for articles | |
| Use of e-resources | |
| *Subtotal* | *180* |
| **Connection to Library** | |
| Relationship with librarians | |
| Librarian involvement with faculty | |
| Trust in Librarians | |
| Involvement in library acquisition (such as on faculty library committee) | |
| *Subtotal* | *125* |
| **Cancellation Project** | |
| Communication with faculty and students | |
| Knowledge of cancellation project | |
| Involvement of faculty in identifying important journals (e.g., filled out faculty survey) | |
| Faculty engagement with cancellation project | |
| Attitude towards journal cancellations | |
| *Subtotal* | *108* |
| **Determining Journal Value** | |
| Field specific considerations | |
| Criteria ranking | |
| Importance of journals to research and/or teaching | |
| Assessment of journal value | |
| Assessment of journal value- impact factor | |
| *Subtotal* | *143* |
| **Cost** | |
| Awareness of cost | |
| Budgetary concerns | |
| *Subtotal* | *98* |
| **Open Access** | |
| Author Processing Charges (APCs) | |
| Publishing in OA (advantages and disadvantages) | |
| Concerns about OA (e.g., predatory journals) | |
| *Subtotal* | *68* |
| **Current Publishing Model** | |
| Attitude towards current publishing model | |
| Alternative ideas on how to reduce journal costs | |
| Difficulty in changing faculty publishing behaviour | |
| *Subtotal* | *140* |
| **Grand total** | **964** |